\documentclass[final,5p,times,twocolumn,authoryear]{elsarticle}

\usepackage{amssymb}
\usepackage{lipsum}

\journal{Journal of Systems and Software (JSS)}

\usepackage{mypreamble}

\definecolor{mygreen}{rgb}{0.0, 0.5, 0.0}

\usepackage{wrapfig}

\usepackage{hyperref}

\begin{document}

\begin{frontmatter}





\title{PyTester: Deep Reinforcement Learning for Text-to-Testcase Generation}

\author[first]{Wannita Takerngsaksiri}
\author[second]{Rujikorn Charakorn}
\author[first]{Chakkrit Tantithamthavorn}
\author[first]{Yuan-Fang Li}

\affiliation[first]{organization={Monash University},country={Australia}}
\affiliation[second]{organization={Vidyasirimedhi Institute of Science and Technology (VISTEC)},country={Thailand}}

\begin{abstract}
Test-driven development (TDD) is a widely-employed software development practice that mandates writing test cases based on a textual description \emph{before} writing the actual code.
While writing test cases is the centerpiece of TDD, it is time-consuming, expensive, and often shunned by developers. 
To address these issues associated with TDD, automated test case generation approaches have recently been investigated.
Such approaches take source code as input, but not the textual description.
Therefore, existing work does not fully support true TDD, as actual code is required to generate test cases. 
In addition, current deep learning-based test case generation approaches are trained with one learning objective, i.e., to generate test cases that are exactly matched with the ground-truth test cases.
However, such approaches may limit the model's ability to generate different yet correct test cases. 
In this paper, we introduce \ours, a Text-to-Testcase generation approach that can
automatically generate syntactically correct, executable, complete, and effective test cases while being aligned with a given textual description.
We evaluate \ours~on the public APPS benchmark dataset, and the results show that our Deep RL approach enables \ours, a small language model, to outperform much larger language models like GPT3.5, StarCoder, and InCoder.
Our findings suggest that future research could consider improving small over large LMs for better resource efficiency by integrating the SE domain knowledge into the design of reinforcement learning architecture.
\end{abstract}






\end{frontmatter}

\section{Introduction}\label{sec:intro}

TDD is an agile software development practice~\citep{beck2001manifesto} that advocates the creation of test cases before code development~\citep{beck2022test,astels2003test}.
The methodology comprises iterative steps of creating test cases, writing code to pass the test cases, and refactoring.
Thus, developers can conceptualize the specification before coding, which increases developers' productivity~\citep{erdogmus2005effectiveness}.
Therefore, writing high-quality test cases is crucial for TDD, yet still time-consuming and complex.

Following TDD practices, developers may focus on writing test cases at different granularity levels (e.g., acceptance, integration, system, and unit test cases) depending on the details of the requirements (e.g., user stories, system description, class description, function description).
However, current research only supports a few scenarios of TDD (e.g., generating acceptance test cases from natural language specifications).
For example, \cite{fischbach2023automatic} and \cite{wang2020automatic} proposed an approach to generate acceptance test cases based on a use case specification.
However, these approaches only generate test cases at the acceptance level for a given use case.
While generating acceptance test cases will greatly help teams to check if all of the use cases are met, it does not guarantee functionality correctness for every method in the software.

Therefore, researchers proposed various automated test case generation approaches to automatically generate test cases at the unit testing level using various techniques, e.g., random techniques~\citep{PachecoLEB2007, pacheco2007randoop}, search-based techniques~\citep{lukasczyk2022pynguin, fraser2011evosuite}, and deep learning techniques~\citep{watson2020atlas, tufano2020athena, alagarsamy2023a3test}.
However, these test case generation approaches often take code as input, which does not apply to TDD, as TDD mandates writing test cases before coding.
Thus, it remains challenging for developers to follow the TDD practices, i.e., writing unit tests based on a given textual description (i.e., the function description). 
Therefore, a novel Text-to-Testcase generation approach is required to help developers write unit tests based on a given textual description, which may improve the overall efficiency of automated software testing.


\emph{In this paper}, we propose \ours, a Text-to-Testcase generation approach for Python using Deep Reinforcement Learning.
Our Deep RL-based model is optimized with Proximal Policy Optimization (PPO) on a reward function that considers three types of feedback, namely, syntax correctness, test executability, and code coverage.
We evaluate our approach on four evaluation aspects, i.e., syntax correctness, passing test case, test completeness measured by code coverage, and test effectiveness measured by mutation score.
Evaluating the alignment of the test cases to a given textual description is indeed challenging.
To address this challenge, we define that a test case is aligned with a given textual description if the test case is executable (i.e., all assertions are passed.) with the ground truth code, assuming that the ground truth code is the correct implementation of the given textual description.
Then, we compare our~\ours~with five strong large language models (LLMs), namely, CodeT5-large\citep{le2022coderl}, GPT3.5\citep{openai2023gpt4}, Copilot\citep{copilot2024}, StarCoder~\citep{li2023starcoder}, and InCoder~\citep{fried2022incoder}.
Through a comprehensive experiment on the APPS benchmark dataset~\citep{hendrycks2021measuring}, we address the following research questions:

\begin{enumerate}[label=\textbf{RQ\arabic*.}, left=0pt]
    \item \textbf{\rqone}\\
    Of the test cases generated by \ours, 99\% are syntactically correct, 84\% are passed (i.e., correct test cases) while achieving a code coverage of 80\% and a mutation score of 61\%, which outperforms all of the studied LLMs.
    This finding demonstrates that a smaller language model like \ours~ can outperform larger language models when carefully designed.
    Importantly, after being optimized with Deep Reinforcement Learning, \ours~outperforms GPT3.5 for all evaluation metrics, highlighting the effectiveness of our Deep RL framework for Text-to-Testcase generation.
    
    \item\textbf{\rqtwo} \\
    The top-3 reward functions for our \ours~are the Syntax+Coverage, Syntax+Executability, and Syntax-only types of feedback.
    Without considering syntax correctness, model performance decreases by 3-13\%, indicating that syntax correctness must be the minimum consideration when designing a reward function for Deep RL in the Text-to-Testcase generation task.
    
    \item \textbf{\rqfour}\\
    As low as 15.59\% of the PyTester-generated test cases are incorrect. Among the incorrect test cases, \ours~mostly encounters an Assertion Error issue for 10.75\%, followed by other types of errors (i.e., IndexError, ValueError, EOFError, SyntaxError).
\end{enumerate}

These results lead us to conclude that the performance of smaller language models like \ours~is not necessarily inferior to that of large language models if carefully designed. 
Our results confirm that Deep Reinforcement Learning substantially contributes to performance improvements.  
Deep RL enables smaller language models, with the smallest model parameter size and the fastest inference times by \emph{at least} one order of magnitude, to outperform much larger language models.
Furthermore, our findings indicate that incorporating test case characteristics into the reward function significantly enhances performance, resulting in the model generating test cases that are 6\% more syntactically correct, 5\% more aligned with the description (i.e., passing test case) with 6\% higher code coverage and 7\% higher mutation score when compared to the GPT3.5 large language model.
These findings emphasize the importance of considering domain knowledge and test case characteristics when designing a Deep RL-based Text-to-Testcase generation approach.

Finally, we acknowledge the importance of the requirement understanding to generate test cases by software engineers. Thus, the main purpose of PyTester is only to assist the TDD process during the test case generation step---\emph{neither to fully automate the TDD process nor replace human involvement.}
Although our PyTester can generate test cases based on a given textual description, it remains essential for software engineers to review and refine these generated test cases to ensure their correctness and completeness.

\textbf{Novelty \& Contributions.} The novelty and contributions of this paper are as follows:
\begin{itemize}
    \revise{R2.1}{
    \item We, \emph{conceptually}, introduce~\ours, a Text-to-Testcase generation approach for a single call function that can automatically generate syntactically correct, executable, complete, and effective test cases while being aligned with a given textual description.
    }
    \item We, \emph{technically}, formulate the Text-to-Testcase generation task as a deep reinforcement learning problem where the reward function is designed to consider various characteristics of test cases, including syntax correctness, test executability, and code coverage.
    \revise{R2.1}{
    \item We, \emph{empirically}, demonstrate that on the APPS dataset, our \ours~model outperforms all of the studied state-of-the-art Large Language Models including GPT3.5, Copilot, StarCoder, and InCoder, while being the smallest language model with the fastest inference time.
    }
\end{itemize}

To support the open science initiatives and increase the verifiability of our study, the replication package of \ours~is available at \url{https://github.com/awsm-research/pytester}.


\section{Background and Motivation}\label{sec:background}

In this section, we describe the background of TDD, discuss existing automated test case generation approaches and a motivating example of ChatGPT for Text-to-Testcase generation.

\subsection{Test-Driven Development (TDD)}

Test-Driven Development (TDD)~\citep{beck2022test, astels2003test} is a software development methodology that involves writing tests for a piece of code before actually implementing the code itself. 
Aligned with the Shift-Left Testing principles and Agile manifesto~\citep{beck2001manifesto, ambler2004object}, TDD is typically carried out in small, incremental steps, with the developer \emph{\textcircled{1} writing a failing test based on textual descriptions, then \textcircled{2} writing the minimum amount of code required to make the test pass, and finally \textcircled{3} refactoring the code as needed}. 
With TDD, developers need to first understand the descriptions and think about the expected behavior of their code, leading to more reliable and maintainable code.
Prior studies also found numerous benefits of TDD, including increasing requirements understanding~\citep{mueller2002experiment}, and producing more clean code with a higher number of test cases~\citep{janzen2008does, erdogmus2005effectiveness}.
However, the process of writing test cases remains manual, time-consuming, and labor-intensive.
This limitation highlights the need for an automated test case generation approach based on textual descriptions, which could potentially accelerate the TDD process and enhance developer productivity.

\subsection{Automated Test Case Generation Approaches}

Various automated test case generation approaches are proposed with different techniques.


\emph{Random-based test case generation} (Randoop), proposed by Pacheco~\ea~\citep{PachecoLEB2007, pacheco2007randoop}, aims to automatically create unit tests for Java classes using feedback-directed random test generation. 
However, the generated test cases are still suboptimal (e.g., miss edge cases or fail to cover specific scenarios) and non-deterministic (e.g., produce different sets of test cases from different runs).

\emph{Search-based test case generation} (e.g., Evosuite~\citep{fraser2011evosuite} for Java and Pynguin~\citep{lukasczyk2022pynguin} for Python) leverages evolutionary search algorithms to guide the search for suitable test cases to produce test suites that achieve high code coverage.
However, prior studies raised concerns that the test cases generated by search-based approaches are often not meaningful ~\citep{almasi2017industrial} and ineffective~\citep{grano2019scented, palomba2016diffusion}.

\emph{Deep learning-based test case generation} (e.g., ATLAS~\citep{watson2020atlas}, Athena~\citep{tufano2020athena}, and A3Test~\citep{alagarsamy2023a3test}) formulated the task as a Neural Machine Translation (NMT) problem (Java Method\textrightarrow Test).
However, such DL-based approaches aim to generate test cases that are exactly matched with the ground truth test cases, without considering the alternative test cases.

\revise{R1.6}{
\emph{NLP-based test case generation} (e.g., Jdoctor~\citep{blasi2018translating}, and TOGA~\citep{dinella2022toga}) applied pattern, lexical, and semantic matching to translate Javadoc comments into test cases (Jdoctor) or a unified transformer-based approach to translate the context of a Java Method into assertion tests (TOGA).
However, these approaches are limited to Java programming language.
}

\textbf{Limitations.} While different test case generation approaches are proposed, none of these approaches are designed for TDD (i.e., taking textual description as input), are not able to generate test cases without code (i.e., mostly they take code as input) and are mostly designed for Java---not Python. 
Therefore, these limitations highlight the need for a Python automated test case generation approach that takes a description as input and can generate alternative test cases beyond the ones that exactly match the ground truth test cases.

\subsection{ChatGPT for Text-to-Testcase Generation: Motivation}\label{subsec:chatgpt}

ChatGPT~\citep{openai2023gpt4} is a powerful AI conversational large language model (LLM) designed for generating natural text based on a given prompt question in a dialogue style.
Recently, researchers found that ChatGPT can perform very well in various software engineering tasks~\citep{hou2023large}, including code generation~\citep{chen2022codet} and test case generation~\citep{schaferempirical,siddiq2023exploring,tang2023chatgpt,xie2023chatunitest,yuan2023no}.
For example, \citet{chen2022codet} proposed CodeT, a code generation approach based on a heuristic ranking process using a dual execution agreement of ChatGPT-generated codes and test cases.
\citet{schaferempirical} proposed TestPilot, a ChatGPT-based test case generation approach for JavaScript based on a given input function.
\citet{siddiq2023exploring} investigated the performance of LLMs for test case generation based on a given Java focal method (i.e., methods to be tested).
Similarly, many recent work~\citep{tang2023chatgpt,xie2023chatunitest,yuan2023no} also investigated the performance of ChatGPT-based test case generation based on a given Java focal method.

While ChatGPT has been widely used for test case generation, existing work only focuses on taking code as input\emph{---not the natural language descriptions as input}.
Therefore, existing test case generation approaches do not support TDD, as TDD mandates writing test cases based on textual descriptions before writing the actual code.
In addition, a systematic literature review by \citet{hou2023large} also confirmed that test case generation approaches to support TDD (i.e., LLM-based Text-to-Testcase generation) remain largely unexplored.
To address this research gap, we first conduct a preliminary analysis to empirically investigate the performance of ChatGPT for Text-to-Testcase generation.





\begin{figure*}[t]
    \centering
    \includegraphics[width=\textwidth]{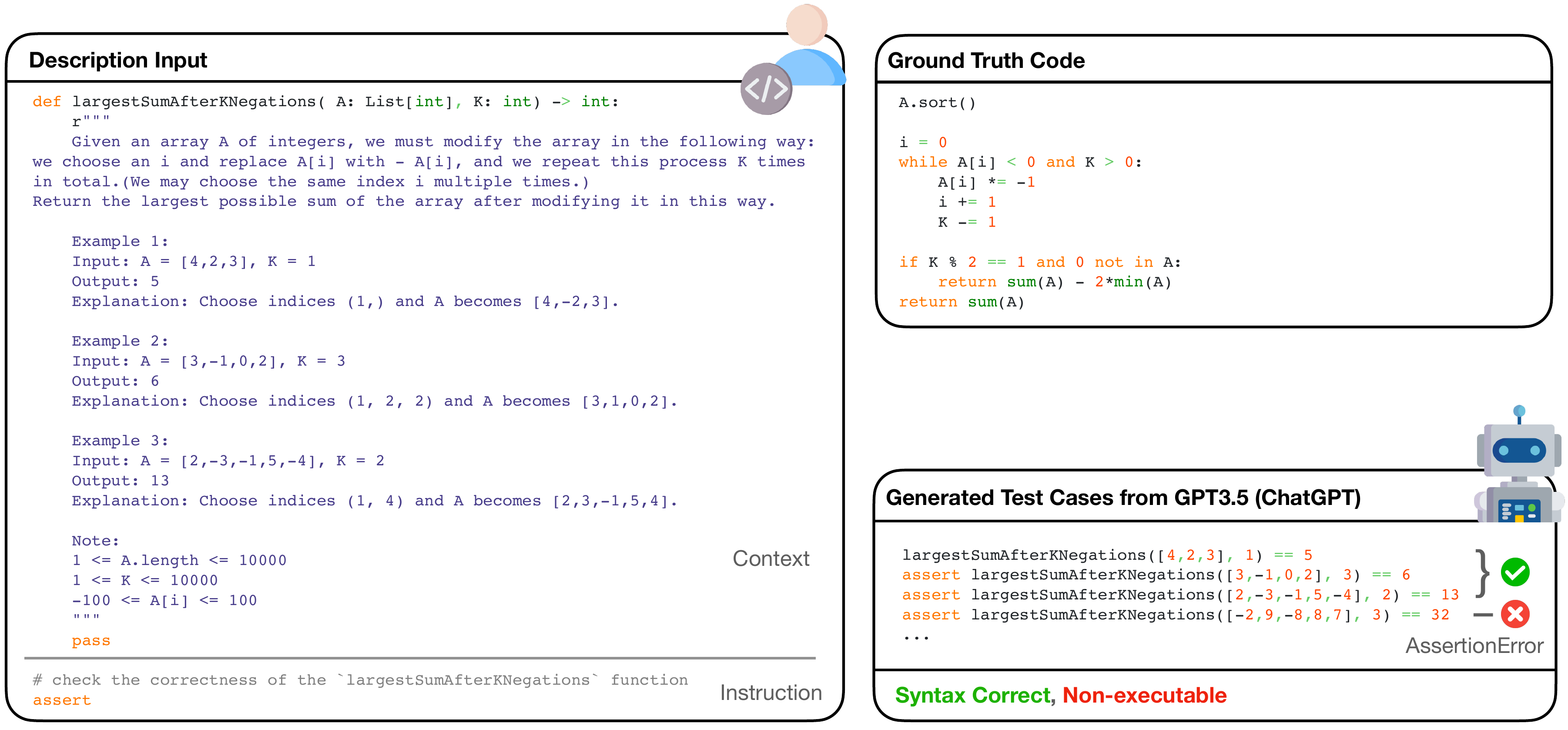}
    \caption{An example of test cases generated by the OpenAI's GPT3.5 model based on a given description input. Syntax correctness is measured by an AST parser, while the test executability is measured by running the test cases against the ground truth code.
    }
    \label{fig:chatgpt}
\end{figure*}


\textbf{An Illustrative Example.} 
We use ChatGPT through the OpenAI API by selecting the recommended \textit{GPT-3.5-turbo} model endpoint.\footnote{The experiment was run in June 2023.}
\revise{R1.5}{
Following \citet{chen2022codet}, 
our prompt includes the context (i.e., external information or additional context that can steer the model to better responses) and the instruction (i.e., a specific task or instruction that the model should perform).
Figure~\ref{fig:chatgpt} (left) presents an example of the input format (i.e., prompt format) where the context refers to the signature of the Python function with a natural language description in the docstring, without the code implementation (Figure~\ref{fig:chatgpt} (top right)), while the instruction refers to a natural language comment ``\emph{\# check the correctness of the `...' function}'' followed by an incompleted \texttt{assert} statement.
}
We randomly select one programming task (Problem \#4479) from the testing set of the APPS dataset~\citep{hendrycks2021measuring} as an illustrative example.
Note that a comprehensive evaluation of ChatGPT is provided in Section~\ref{sec:exp_result} where we found that our \ours~performs better.
For the given Problem \#4479, we reformat the data structure into a triplet of $<$\emph{function signature + description text}, \emph{code implementation}, \emph{test cases}$>$.
Then, the \emph{function signature + description text} is input into ChatGPT to generate test cases, while \emph{code implementation} is only used to evaluate if the generated test cases can be run against the ground truth code.

\textbf{Observation-1: ChatGPT can generate syntactically correct test cases, yet some of the generated test cases cannot be executed (i.e., failed test cases).} 
In reality, when using ChatGPT to generate test cases based on a given description, it becomes challenging to verify if the generated test cases are correct and aligned with the given description or not.
To address this challenge, assuming that the ground truth code is the correct implementation, we execute the test cases against the ground truth code to verify if the ChatGPT-generated test cases are executable or not.
Figure~\ref{fig:chatgpt} (bottom right) shows that four test cases generated by ChatGPT are syntactically correct, but the fourth test case is not executable. 
In particular, the fourth test case is generated with the test input of \texttt{`([-2,9,-8,8,7], 3)'}.
Thus, the expected output should be \texttt{`30'}, but incorrectly generated as \texttt{`32'}.
Therefore, this test case will raise an \emph{AssertionError}.
Such syntactically correct, yet non-executable test cases indicate that the test case is still not aligned with the given descriptions.
This observation demonstrates that ChatGPT is able to generate syntactically correct test cases, yet fails to generate executable ones (i.e., failed test cases).

\textbf{Observation-2: When ChatGPT generates non-executable test cases, it becomes challenging to evaluate the completeness and effectiveness of the test cases.} 
Intuitively, effective test cases should exercise all lines of code in the program (i.e., achieving high code coverage), and be able to discover bugs/defects when the program is slightly changed (i.e., the same set of test cases is able to catch the introduced bugs).
Code coverage is widely used to measure test completeness ~\citep{pacheco2007randoop, fraser2011evosuite, schaferempirical, siddiq2023exploring, tufano2020athena, alagarsamy2023a3test, tang2023chatgpt}.
Thus, a higher code coverage percentage often increases the confidence that the software behaves as intended.
However, studies~\citep{hemmati2015effective, cai2005effect} argue that code coverage only measures the extent to which the test cases exercise the lines of code in the program, but not the effectiveness of the test cases.
Therefore, mutation testing is proposed to evaluate the effectiveness of the test cases~\citep{cai2005effect, madeyski2010impact, jia2010analysis, papadakis2019mutation}.
Mutation testing aims to introduce intentional faults (mutations) into the code and check if the existing test cases can detect these faults (mutants)---\emph{referring to the fault detection capability of the test cases}.
Thus, the mutation score is measured as a percentage based on the number of mutants that are killed (i.e.\ detected by the test cases) versus the total number of mutants introduced.
Nonetheless, despite the benefits of code coverage and the extensive research on mutation testing in the literature, another crucial factor is the test case executability.
When ChatGPT generates non-executable test cases, such critical metrics (e.g., code coverage and mutation score) are unable to be assessed.
Therefore, the models should consider these domain knowledge metrics during training – which is largely overlooked in the existing ChatGPT-based approaches.
For instance, although \citep{chen2022codet} proposed a method to re-rank generated code and test cases based on execution feedback, it does not take into account the test case characteristic.

\begin{tcolorbox}
\textbf{Summary.}
TDD offers numerous benefits to software development, yet is time-consuming and expensive. 
Existing test case generation approaches do not yet support TDD practices (i.e., code, \emph{not textual description}, is treated as input).
Our motivating analysis shows that ChatGPT, one of the most powerful language models, still generates non-executable, incomplete, and ineffective test cases, hindering the adoption of ChatGPT and other LLMs for Text-to-Testcase generation in practice.
To support TDD practices, this observation highlights the need for a novel automated test case generation approach that can generate syntactically correct, executable, complete, and effective test cases.
\end{tcolorbox}

\section{\ours}\label{sec:approach}




%

\textbf{Aim.} We introduce \ours, a Text-to-Testcase generation approach that aims to automatically generate syntactically correct, executable, complete, and effective test cases based on a given natural language description. 
We formulate the Text-to-Testcase generation task as a Reinforcement Learning (RL) problem due to the following reasons:

\begin{itemize}
    \item \textbf{To increase models' ability to generate alternative test cases.} 
    Traditionally, supervised finetuning approaches are trained to generate test cases that are textually and exactly matched with the ground truth test cases.
    Thus, these models can only generate test cases based on what the model had seen in the training data, limiting the model's ability to generate test cases outside of the domain (i.e., \emph{out-of-distribution (OOD) problems}~\citep{shen2021towards}).
    In fact, the ground truth test cases may not be representative of all of the possible test cases.
    Thus, alternative test cases with different test inputs that still satisfy the given description should also be considered as valid.
    \item \textbf{To incorporate test case characteristics into the feedback loop.} 
    Traditionally, the generated test cases are evaluated only on their similarity with the ground truth test cases.
    However, in real-world practices, developers normally design test cases by ensuring test cases are syntactically correct, executable, and complete.
    Such test case characteristics are critically important, yet remain largely disregarded by the existing DL-based test case generation approaches.
\end{itemize}



\begin{figure}
  \begin{center}
    \includegraphics[width=\linewidth]{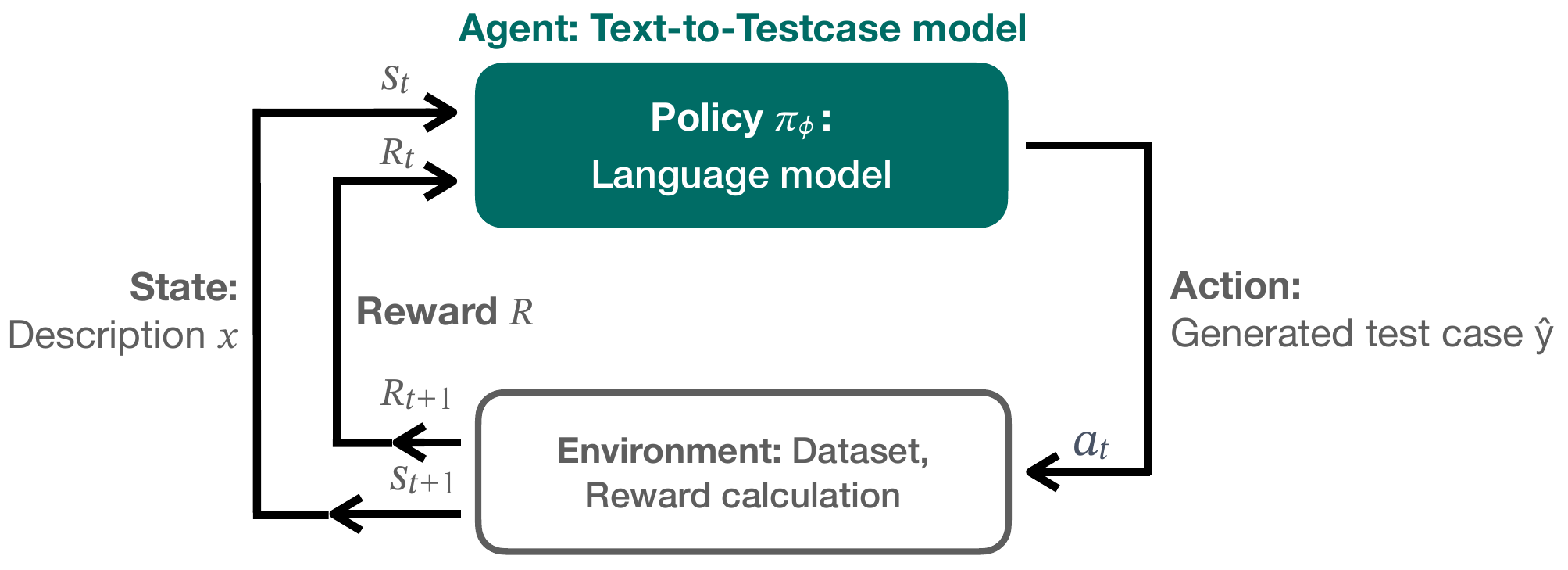}
  \end{center}
  \caption{The formulation of Text-to-Testcase generation as a Reinforcement Learning problem.}
  \label{fig:basic_rl}
\end{figure}

\ours~leverages a Deep Reinforcement Learning (Deep RL) framework, which is a combination of deep learning and reinforcement learning principles.
In the context of the Text-to-Testcase generation, the goal of RL is to enable the agent (i.e.\ the Text-to-Testcase generation model) and its policy (the brain of the agent) to learn by interacting with the environment (through trial and error) and receiving rewards (negative or positive) as feedback for performing actions (i.e.\ generating test cases).
This formulation enables \ours~to (1) learn autonomously through interaction with their environment without ground truth test cases, allowing the models to generate alternative test cases and (2) learn how to generate syntactically correct, executable, complete, and effective test cases while being aligned with a given description through a carefully-crafted reward function rather than a basic loss function used in traditional finetuning models.

\subsection{Our Reinforcement Learning Framework for Text-to-Testcase Generation}\label{subsec:RL}

We represent the reinforcement learning problem as a Markov Decision Process (MDP) which decomposes scenarios as a series of states, connected by actions and associated with a specific reward. 
Following Figure~\ref{fig:basic_rl}, we define and explain each of the components below.

\textbf{Language Model:} Given the natural language \emph{description} $x$ and a \emph{testcase} $y$, the Text-to-Testcase generation language model is defined as $P(y|x) = \prod_{i=0}^m p_\theta(y_i|x, y_{0:i-1})$

\textbf{Dataset:} We define a training dataset $\mathcal{D}$ of size $M$, which contains triplets of $<$\emph{description}, \emph{code}, \emph{test cases}$>$ as $\mathcal{D} = \{(x^i,c^i,y^i)\}_{i=1}^{i=M}$.
The description $x$ consists of a signature of the Python function, the natural language description included in the docstring, the hidden code implementation (\texttt{pass}), the instruction prompt as a natural language comment ``\texttt{\# check the correctness of the `...' function}'', and an incompleted assert statement ``\texttt{assert}''.
On the other hand, the ground truth test case $y$ consists of an assertion statement.


\textbf{Agent/Policy:} Our policy $\pi_\phi$ is defined as a language model $p_\theta$, training from dataset $\mathcal{D}$.
The policy outputs the probability of all vocabularies conditioned on the current state: $\pi_\phi(\cdot|s_t)$.

\textbf{State:} A state $s_t = (x,\hat{y}_{0:t-1})$ is defined as a concatenation of the \textbf{textual description} $x$ and the \textbf{test case} tokens $\hat{y}_{0:t-1}$ generated so far.

\textbf{Action:} An action $a_t = \hat{y}_t \sim \pi_\phi(\cdot|s_t)$ is a token sampled from the policy. The action is then used as a continuation of the current state by the \textbf{transition function}, $T$, to produce the next state, $s_{t+1}$. Specifically, the next state is the concatenation of the action and the current state: $s_{t+1} = T(s_t, a_t) = (x, \hat{y}_t)$, where $\hat{y}_t=a_t\oplus\hat{y}_{0:t-1}$.

With the Markov Decision Process (MDP), our \ours{} agent is learned as an episodic task.
Generally, an episode is a sequence of interactions between the agent and environment starting from an initial state $s_0$ and ending at a terminal state where the agent will receive the reward and update the policy.
At the beginning of each episode, the \emph{initial state} will start from a description $x$ sampled from the Dataset $\mathcal{D}$, denoted as $s_0 = x \sim \mathcal{D}_x$. 
Then, our policy $\pi_\phi$ takes a description $x$ as input to generate a test case $\hat{y}$ that is syntactically correct, executable, and complete. 
Technically, the policy $\pi_\phi$ will sequentially generate a sequence of tokens $\hat{y}_i$ for the test case $\hat{y}$ until reaching the \emph{end of the trajectory} (i.e., after generating the \texttt{EOS} token or reaching the maximum generation length).
Then, the generated test cases $\hat{y}$ are input into the interactive environment to receive a reward via a Reward Calculation function $R(x,\hat{y})$, which returns a cumulative  \textbf{reward point}.
The reward will be provided if the generated test cases are syntactically correct, executable against the ground truth code (i.e., aligned with the descriptions), and complete (i.e., achieving high code coverage).
On the other hand, the penalty will be provided if the generated test cases are syntactically incorrect and non-executable (e.g., runtime errors like incorrect method names and assertion errors like wrong test inputs).
Finally, the policy $\pi_\phi$ is learned over time in order to maximize a reward function written as:

\begin{align}
\mathcal{J}(\pi_\phi) = \mathbb{E}_{x \sim \mathcal{D}_x, \hat{y} \sim \pi_\phi(\cdot|x)}R(x,\hat{y})
\end{align}

\begin{figure*}[t]
    \centering
    \includegraphics[width=\textwidth]{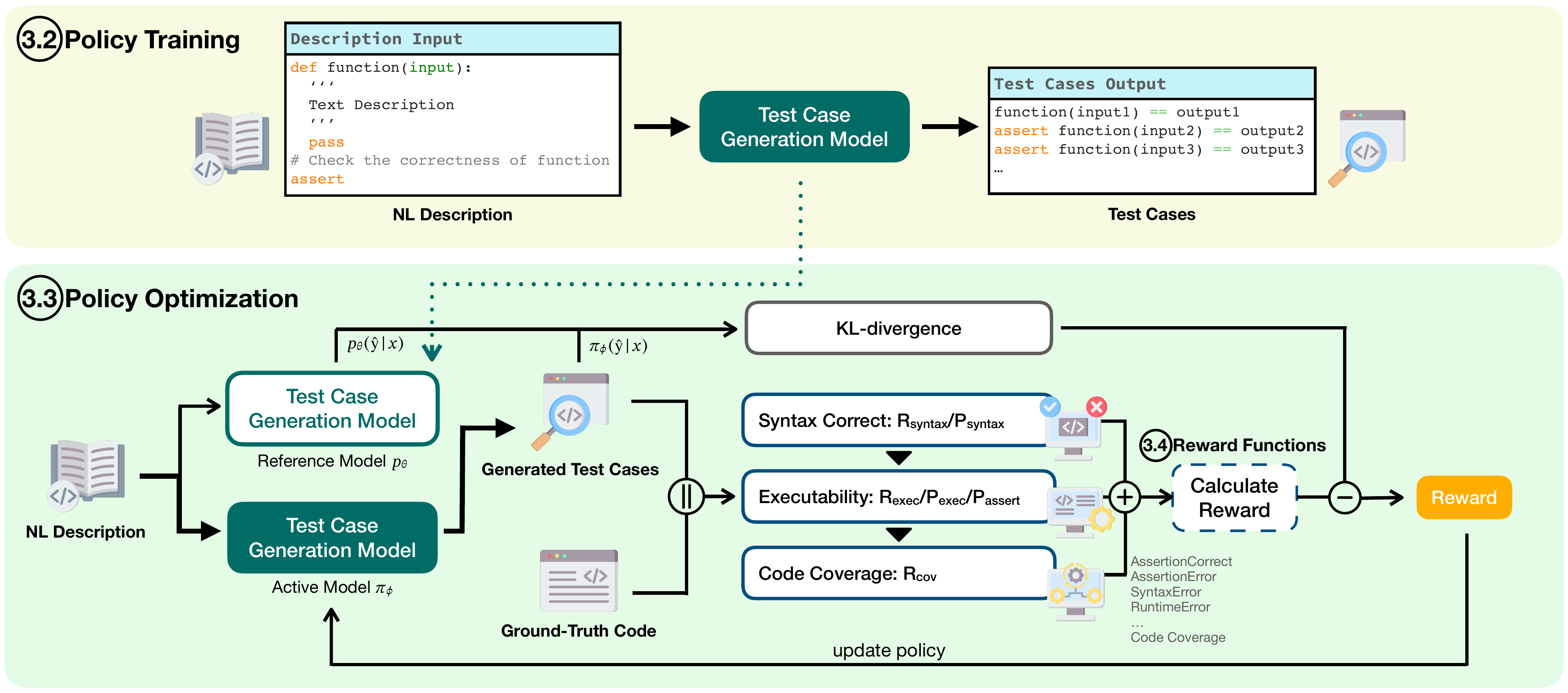}
    \caption{An overview of our 
    Deep Reinforcement Learning framework for the Text-to-Testcase Generation task (called \ours).
    $||$ denotes a concatenation, $+$ denotes an addition, and $-$ denotes a subtraction.} 
    \label{fig:overview}
\end{figure*}

\textbf{Overview.} Figure~\ref{fig:overview} presents an overview of our \ours~framework.
Our framework consists of two steps: policy training and policy optimization.
In the \emph{policy training} step, we aim to build the agent (i.e., the Text-to-Testcase Generation model) where its policy interacts with the environment (i.e., the input dataset) and learns to make decisions (i.e., generate test cases) over time.
However, currently, the agent is still suboptimal as it is trained to generate the same test cases as the ground truths in a supervised manner.
Thus, the \emph{policy optimization} step aims to further optimize the agent via a reward function through Deep Reinforcement Learning.
To do so, given an environment (i.e., a triplet of $<$\emph{description}, \emph{code}, \emph{test cases}$>$ in the training dataset), the agent will take the description as input into the model in order to perform an action (i.e., generate test cases), where the test cases will be evaluated via a reward function.
Our reward function is designed to incorporate three aspects of test case characteristics, namely syntax correctness, test executability, and test completeness.
The agent is optimized with Proximal Policy Optimization (PPO)~\citep{schulman2017proximal} using our reward function, and stabilized with KL-divergence~\citep{kullback1951information}.
Below, we describe the technical details of policy training,  policy optimization, and our reward function.

\subsection{Policy Training}

The ultimate goal of the policy training is to build a language model with some preliminary knowledge of the relationship between the description and its associated test cases.
Therefore, we adopt the \emph{default CodeT5-large} as our language model~\citep{le2022coderl,raffel2020exploring}.
To ensure that the model performs well on a target task (i.e., Text-to-Testcase generation), we finetune our language model with the training dataset.
Given a training dataset $\mathcal{D} = \{(x^i,c^i,y^i)\}_{i=1}^{i=M}$, we finetune the language model to learn the mapping between the description $x$ and its corresponding test cases $y$  (not the ground truth code $c$) in order to minimize the multi-class cross-entropy loss $\mathcal{L}_\text{FT} = - \mathbb{E}_{\substack{(x,y) \sim \mathcal{D}\\n \in \{0,..., N\}}} \text{log } p_\theta(y_n|x,y_{0:n-1})$, where $N$ is the maximum length of the test cases.
Then, the loss is optimized using an Adam optimizer~\citep{kingma2014adam} (the hyperparameters settings are provided in the Appendix).

\subsection{Policy Optimization}



Next, we further update the \ours~policy with a reward function through reinforcement learning. 
At the initial state, we initiate two finetuned language models from the policy training step: a language model policy $\pi_\phi$ as our active model (i.e., to-be trained model) and a language model $p_\theta$ as a reference model (i.e., the frozen model that is not updated during the policy optimization step).
We optimize our policy using PPO and KL-divergence as follows.   

\textbf{Proximal Policy Optimization (PPO)}~\citep{schulman2017proximal}. 
We use PPO to improve the training stability of the policy optimization process. 
Formally, PPO optimizes the following surrogate objective:
\begin{align}
    \mathcal{J}_\text{surr}(\pi_\phi) = \text{min} &(r A^{\pi_{\phi_\text{old}}}(x,\hat{y}), \text{clip}(r, 1-\epsilon, 1+\epsilon) A^{\pi_{\phi_\text{old}}}(x,\hat{y})), \\
    \text{where} \quad   r &= \frac{\pi_\phi(\hat{y}|x)}{\pi_{\phi_\text{old}}(\hat{y}|x)},
\end{align}
in which $\epsilon$ is a small positive scalar indicating how much the policy can be updated and $\pi_{\phi_\text{old}}$ is the policy that collects the data. 
$A^\pi(x,\hat{y})$ is the advantage function of a policy $\pi$ given a state-action (i.e., input text description - output test cases) pair $(x,\hat{y})$, and is computed by the Generalized Advantage Estimator (GAE)~\citep{schulman2015high}.

\textbf{KL-divergence}~\citep{kullback1951information}. 
Following best practices \citep{ouyang2022training,stiennon2020learning,ziegler2019fine}, we utilize the Kullback-Leibler (KL) divergence regularization~\citep{kullback1951information} to stabilize the training process.
KL-divergence is a penalty to prevent our active model $\pi_\phi$ from exploring too differently from the reference model $p_\theta$ (i.e., the finetuned model). 
Concretely, we define the KL-divergence as $KL_{penalty}$ and the reward function $R$ as
\begin{align}
R(x,\hat{y}) = R_\text{task} - KL_\text{penalty} = R_\text{task} - \beta \text{log} \frac{\pi_\phi(\hat{y}|x)}{p_\theta(\hat{y}|x)},
\end{align}
where $R_{task}$ is the summation of the rewards and $\beta > 0$ is a hyperparameter controlling the strength of the regularization.


\subsection{Reward Functions}~\label{subsec:rewards}

Once the test cases $\hat{y}$ are generated from the model, the generated test cases are evaluated against our reward function.
Our reward function aims to consider three types of feedback, namely, syntax correctness, test executability, and test completeness. 
In general, each reward function will provide a reward point for positive feedback or a penalty point for negative feedback.
Our policy is optimized based on the following reward function:

\begin{align}
R_{task} =  R_{syntax} +  R_{exec} +  R_{cc} - P_{syntax} - P_{exec} - P_{assert}
\end{align}

Below, we define the three types of feedback as follows:

\textbf{Syntax Correctness Feedback} concerns the syntactically correctness of the generated test cases $\hat{y}$. 
To determine if the generated test cases are syntactically correct, we use the Python \texttt{ast} library\footnote{https://docs.python.org/3/library/ast.html} to parse the generated test cases $\hat{y}$.
We define Condition $\mathds{1}_{syntax(\hat{y})} = 1$ if $\hat{y}$ is syntactically correct (i.e., successfully parsed by AST). 
Then, the feedback is provided to the policy based on the following reward/penalty function given reward/penalty point ($r_{syntax}, p_{syntax}$):
\begin{align}
    R_{syntax} &= \mathds{1}_{syntax}r_{syntax}, \\
    P_{syntax} &= (1-\mathds{1}_{syntax})p_{syntax},
\end{align}







\textbf{Test Executability Feedback} concerns the functional correctness of the generated test cases $\hat{y}$ given the ground truth code $c$.
To determine if the generated test cases are executable and functional correct to the ground truth code, we use the built-in function \texttt{exec}.\footnote{https://docs.python.org/3/library/functions.html\#exec}
Specifically, we decompose the executability feedback into two conditions: Execution Condition ($\mathds{1}_{exec(c,\hat{y})}$) and Assertion Condition ($\mathds{1}_{assert(c,\hat{y})}$). We define Condition $\mathds{1}_{exec(c,\hat{y})} = 1$ if the execution of ($c,\hat{y}$) does not raise a \texttt{RuntimeError} (i.e., the test cases are executable) and Condition $\mathds{1}_{assert(c,\hat{y})} = 1$ if the execution does not raise an \texttt{AssertionError} (i.e., execution passes all the generated assertion test cases).
Then, the feedback is provided to the policy based on the following reward/penalty function given reward/penalty point ($r_{exec}, p_{exec}, p_{assert}$):
\begin{align}
    R_{exec} &= \mathds{1}_{syntax}\mathds{1}_{exec}\mathds{1}_{assert}r_{exec}, \\
    P_{exec} &= \mathds{1}_{syntax}(1-\mathds{1}_{exec})p_{exec},\\
    P_{assert} &= \mathds{1}_{syntax}\mathds{1}_{exec}(1-\mathds{1}_{assert})p_{assert}
\end{align}

\textbf{Code Coverage Feedback} concerns the percentage of the line of code (LOC) in ground truth code $c$ tested by generated test cases $\hat{y}$.
To determine the percentage, we use \texttt{coverage.py}\footnote{https://coverage.readthedocs.io/en/7.2.7/} to calculate the code coverage, denoted by $CodeCoverage(c, \hat{y})$, which has a value range of $[0,100]$. 
Then, the feedback is provided to the policy based on the following reward function given a normalized $CodeCoverage(c, \hat{y})$ as reward point ($r_{cov}$):
\begin{align}
R_{cov} = \mathds{1}_{syntax}\mathds{1}_{exec}\mathds{1}_{assert}r_{cov}
\end{align}
\section{Experimental Setup}\label{sec:exp_setup}

In this section, we present the motivation of our three research questions and describe our experimental setup for evaluating our \ours~approach and the baselines.



\revise{R1.6}{
\begin{enumerate}[label=\textbf{RQ\arabic*.}, left=0pt]
    \item \textbf{\rqone} \\
    \emph{Motivation.} Prior studies found that large language models (e.g., GPT3.5) tend to outperform smaller language models for various SE tasks \citep{openai2023gpt4, roziere2023code}.
    However, little is known about whether our \ours~which is based on a smaller language model like CodeT5 will outperform large language models.

    \item \textbf{\rqtwo} \\
    \emph{Motivation.} The reward function plays an important role in providing a positive/negative signal to the agent if the action satisfies the goal.
    There exist multiple variations of the reward functions (e.g., considering either one or multiple types of feedback).
    Yet, little is known about which types of feedback are most important. 
    
    \item \textbf{\rqfour} \\
    \emph{Motivation.} Our \ours~approach may generate both correct and incorrect test cases.
    Thus, a manual analysis of the incorrectly generated test cases may offer opportunities for future researchers to consider improving the performance of our approach.
    Yet, little is known about what are the most common types of errors that are found in the incorrectly generated test cases by \ours. 
\end{enumerate}
}

\subsection{Studied Dataset}

\textbf{The APPS benchmark dataset}~\citep{hendrycks2021measuring} is used to evaluate our approach and compare it with the baseline approaches. 
The APPS benchmark dataset consists of 10,000 programming tasks with 131,777 associated test cases collected from Codewars, AtCoder, Kattis, and Codeforces.
The dataset consists of three difficulty levels of the programming tasks, including the introductory level, the interview level, and the coding competition level.
Each of the programming tasks consists of triplets of $<$\emph{a description, an associated code, and associated test cases}$>$.
Similar to prior studies~\citep{hendrycks2021measuring}, we use the same dataset splitting of 5,000 samples for training and 5,000 samples for testing.

\revise{R1.3}{
Notably, in this study, only the APPS benchmark dataset is suitable for training our Text-to-Testcase generation task, as our \ours~model requires a large enough size of the dataset for policy training.
Nevertheless, to broaden our evaluation, we provide additional analysis of PyTester in a zero-shot setting using \textbf{MBPP}~\citep{austin2021mbpp} and \textbf{HumanEval}~\citep{chen2021humaneval} datasets.
More details can be found in the discussion section (see Section~\ref{sec:discussion}).}

\subsection{Data Preprocessing}

The quality of the dataset may have an impact on the performance of \ours. 
Thus, we carefully craft the dataset for model training through our data preprocessing as follows.

\textbf{Data Reformatting.} 
After manual analysis, we found that, in the APPS dataset, different programming tasks have different approaches to test case execution.
In particular, some programming tasks are tested with an assert statement format (\texttt{assert add(1,1) == 2}), while some are tested with a standard input-output format (\texttt{python add.py 1 1 2}).
Thus, it becomes challenging for a DL model to automatically handle such inconsistency.
In particular, for such programming tasks with the standard input-output format, the method name does not exist and the assert statements are missing.
To address this challenge, for programming tasks written in a standard input-output format, we transform them by giving a function name to the programming task, then adding a test method (e.g., \texttt{def test\_main()}) to be called by assert statements (see an example in Figure~\ref{fig:rq1_example}).
Finally, all of the programming tasks in the dataset will have a consistent format of assertion calls.






\textbf{Data Filtering.}
In addition, we found that not all samples in the dataset are of high quality.
Common issues include missing test cases, missing code solutions, and non-executable test cases.
Since the goal of \ours~is to generate test cases that are aligned with the natural language description, our selection criteria are defined as the test cases that must exist and be executable against the code solution. 
After the data filtering step, our training dataset consists of 3,401 samples and our testing dataset consists of 3,259 samples.
During the model training step, we use 10\% of the training dataset for validation, while the rest of the training dataset is used for model training.


\subsection{Evaluation Metrics}

Since the goal of \ours~is to generate syntactically correct, executable, complete, and effective test cases that are aligned with the description, our evaluation will focus on the following four aspects: syntax correctness, passing test case, code coverage, and mutation score.

\begin{itemize}
    \item \textbf{Syntax Correctness} measures the percentage of the generated test cases that are syntactically correct. The syntax correctness is defined by the \texttt{ast} Python library.\footnote{https://docs.python.org/3/library/ast.html}
    The given test case is considered syntactically correct if the AST can be successfully run. 

    \item \textbf{Passing Test Case} measures the percentage of the passing generated test cases to the ground truth code.
    Since the ground truth code is the correct implementation of the natural language description, any test cases that can be successfully executed against the ground truth code are considered passed (i.e., call the correct functions to be tested with the right test inputs and the expected outputs).
    Thus, we measure the passing test case by executing the concatenation of the ground truth code with the generated test cases using the \texttt{exec} built-in function.\footnote{https://docs.python.org/3/library/functions.html\#exec}

    
    \item \textbf{Code Coverage} measures the completeness of the generated test cases for a given natural language description.
    Ideally, in addition to the passing test case, the test cases must be complete (i.e. executing all lines of the ground truth code solution).
    Thus, we measure the line-level code coverage of the generated test cases using the \texttt{coverage.py} Python library.\footnote{https://coverage.readthedocs.io/en/7.2.7/}
    
    \item \textbf{Mutation Score}
    measures the effectiveness of the generated test cases.
    High-quality test cases must be able to detect bugs if the code is slightly changed.
    Mutation testing aims to introduce intentional faults (mutations) into the code and check if the existing test cases can detect these faults (mutations).
    Then, the mutation score is measured as a percentage based on the number of mutations that were killed (detected) versus the total number of mutations introduced.
    \revise{R1.4}{
    We use the \texttt{MutPy} Python library\footnote{https://pypi.org/project/MutPy/} to measure a mutation score with a timeout of 30 seconds per sample.
    We set the configuration to default with \texttt{PyTest}\footnote{https://docs.pytest.org/en/latest/} as a test runner.
    The mutation operators are listed in Table~\ref{tab:mut_op}.
    }
\end{itemize}
\begin{table}[h]
    \color{black}
    \centering
    \caption{The operators used in our mutation testing.}
    \begin{tabular}{l}
    Mutation Testing Operator                       \\
    \hline
    AOD - arithmetic operator deletion              \\
    AOR - arithmetic operator replacement           \\
    ASR - assignment operator replacement           \\
    BCR - break continue replacement                \\
    COD - conditional operator deletion             \\
    COI - conditional operator insertion            \\
    CRP - constant replacement                      \\
    DDL - decorator deletion                        \\
    EHD - exception handler deletion                \\
    EXS - exception swallowing                      \\
    IHD - hiding variable deletion                  \\
    IOD - overriding method deletion                \\
    IOP - overridden method calling position change \\
    LCR - logical connector replacement             \\
    LOD - logical operator deletion                 \\
    LOR - logical operator replacement              \\
    ROR - relational operator replacement           \\
    SCD - super calling deletion                    \\
    SCI - super calling insert                      \\
    SIR - slice index remove                       
    \end{tabular}
    \label{tab:mut_op}
\end{table}
    



Different from prior studies that aim to generate test cases that are textually and exactly matched with the ground truth test cases, we neither use such an exact match nor a code similarity measure (e.g., a BLEU score) in this study.


\subsection{Evaluation Setting}
Since the number of test cases generated from each model varies, we provide 3 evaluation settings: Raw, \emph{n}-test, and Filtered by Syntax Correctness, to ensure a fair comparison.
\begin{itemize}
    \item \textbf{Raw} evaluates the generated test cases directly from the model without any post-processing.
    In this setting, the test cases are assessed as a whole set. 
    If any single test case fails, the entire output is considered incorrect.
    
    \item \textbf{\emph{n}-test} evaluates the generated test cases by limiting the size to \emph{n} test cases. 
    We use \emph{n}=3 (3-test) and \emph{n}=5 (5-test).
    
    \item \textbf{Filtered by Syntax Correctness}  evaluates the generated test cases after filtering out the incorrect syntax of the test cases.
    First, we split multiple assertions into individual tests, then exclude the syntactically incorrect test cases before evaluation.
\end{itemize}

\subsection{Baselines}\label{baselines}

Since the existing test case generation approaches (discussed in Section~\ref{sec:background}) are not designed for TDD (i.e. do not take a description as input), it becomes challenging to fairly compare our \ours~with the existing ones.
Since prior studies demonstrated that Large Language Models (LLMs) could be a viable solution for various SE tasks, we consider the following five baseline models.
Note that if available, we use beam search with a size of 5 as the decoding method. Otherwise, we conduct the experiments five times per data point and report the mean and standard deviation.

\begin{itemize}
    \item \textbf{CodeT5-large} (770M parameters)~\citep{le2022coderl} is a transformer language model pre-trained on six programming languages of CodeSearchNet~\citep{husain2019codesearchnet} with the masked span prediction task~\citep{CodeT52021}.
    
    \item \textbf{InCoder} (6.7B parameters)~\citep{fried2022incoder} is a  transformer-based LLM pre-trained on a total of 159GB public code with permissive open source licenses from GitHub and GitLab, in which 52GB are Python, and a total of 57GB of text content are from StackOverflow.
    
    \item \textbf{StarCoder} (15.5B parameters)~\citep{li2023starcoder} is a  transformer-based LLM pre-trained on 80 different programming languages and natural languages such as GitHub issues, commits, and notebooks. 
    We use the finetuned version which is trained with 35B Python tokens.
    
    \item \textbf{GPT-3.5} (\textgreater{175B} parameters\footnote{GPT-3.5 model size is not reported publicly. The previous version (GPT-3 model\citep{brown2020language}) has a model parameter size of 175B.})~\citep{openai2023gpt4} is one of the most powerful state-of-the-art LLMs developed by OpenAI, which is a transformer-based LLM pre-trained with the next token prediction task on the internet and the third-party licensed data, then finetuned using Reinforcement Learning from Human Feedback (RLHF)~\citep{christiano2017deep}.
    We chose the recommended \textit{gpt-3.5-turbo} endpoint, which was accessed in June 2023.

    \item \textbf{Copilot} (\textgreater{12B} parameters\footnote{Copilot model size is not reported publicly. The previous version (Codex model\citep{chen2021humaneval}) has a model parameter size of 12B.})~\citep{copilot2024} is another one of the most powerful state-of-the-art LLMs from GitHub.
    It was powered by the OpenAI Codex Model and introduced as an ``AI pair programmer" in Visual Studio Code.
    We include Copilot in this study to reflect the performance of a tool in a realistic coding scenario. 
    Access to Copilot was obtained through the following repository.\footnote{https://github.com/B00TK1D/copilot-api/tree/main}
    \revise{R1.2}{
    \item \textbf{Pynguin}~\citep{lukasczyk2022pynguin} is a search-based test case generation approach that produces test cases given a source code.
    We include Pynguin in this study to reflect the performance of a traditional Code-to-Testcase tool.
    Notably, due to a technical issue with IO functions, Pynguin is incompatible with the APPS dataset. Nonetheless, we include Pynguin in the discussion section (see Section~\ref{sec:discussion}), where we assess our \ours~in the zero-shot setting on the MBPP and HumanEval datasets.
    }
\end{itemize}

\subsection{Model Experimentation}


The performance of our \ours~model heavily relies on the design of our reward function.
Thus, considering three different types of feedback, we perform an experiment with the following seven variations:
\begin{itemize}
    \item Syntax-only ($p_{syntax}$=-2, $r_{syntax}$=2)
    \item Executability-only ($p_{exec}$=-2, $p_{assert}$=-0.3, $r_{exec}$=2)
    \item Coverage-only ($p_{exec}$=-2, $r_{cov}$=$norm([0, 2])$)
    \item Syntax+Executability ($p_{syntax}$=-2, $p_{exec}$=-1, $p_{assert}$=-0.3, $r_{exec}$=2)
    \item Syntax+Coverage ($p_{syntax}$=-2, $r_{cov}$=$norm([0, 2])$) 
    \item Executability+Coverage ($p_{exec}$=-2, $p_{assert}$=-0.3, $r_{exec}$=2, $r_{cov}$=$norm([0, 2])$)
    \item Syntax+Executability+Coverage ($p_{syntax}$=-2, $p_{exec}$=-1, $p_{assert}$=-0.3, $r_{exec}$=2, $r_{cov}$=$norm([0, 2])$)
\end{itemize}

\noindent where $r$ is the reward point, and $p$ is a penalty point, or 0 if not specified. 
For example, when considering 'Syntax-only' as a reward function, a reward point of 2 is given if the generated test cases are syntactically correct, otherwise, a penalty point of -2 is given.

We run our experiment on two NVIDIA GeForce RTX 3090 GPUs with 24 GB vRAM, an Intel(R) Core(TM) i9-9980XE CPU @ 3.00GHz with 36 core processors, and 64G RAM.
The hyperparameter settings for the policy training and the policy optimization are reported in Table~\ref{tab:ft_hparams} and Table~\ref{tab:rl_hparams}, respectively.

In total, we experimented with 7 variations.
During the training phase, we set the model to learn to generate multiple assert statements (without the length limit). 
During the inference phase, the test case generation for each model uses a beam search of 5 and each model is evaluated using the following measure, called TestQuality:

\begin{equation}
\mathrm{TestQuality} = \frac{(2 \times Code Coverage \times Mutation Score)}{(Code Coverage + Mutation Score)}
\end{equation}

Ideally, high-quality test cases must execute all lines of code in the function (i.e., measured via code coverage), while being able to detect bugs if the code is slightly changed (i.e., measured via mutation score).
We define the Test Quality score as a harmonic mean of the code coverage and the mutation score.
After the comprehensive experiment, we find that \textbf{the \ours~model performs best when it is trained to generate multiple assert statements with a reward function considering both Syntax and Coverage feedback,} which we used as a reference setting for our \ours~model for the remainder of the paper.

\begin{table*}[h]
\parbox{.48\linewidth}{
\centering
    \caption{The hyperparameter settings of \ours~during the policy training step.}
    \begin{tabular}{c|c}
        \textbf{Hyperparameter} & \textbf{Value} \\
        \hline
        Learning rate & 2e-5 \\
        Warmup steps & 1000 \\
        Weight decay & 0.01 \\ 
        Batch size & 2 \\
        Gradient accumulation steps & 4 \\
        Learning rate scheduler type & inverse\_sqrt \\ 
        Epoch & 20 \\
    \end{tabular}
    \label{tab:ft_hparams}
}
\hfill
\parbox{.48\linewidth}{
\centering
    \caption{The hyperparameter settings of \ours~during the policy optimization step.}
    \begin{tabular}{c|c}
        \textbf{Hyperparameter} & \textbf{Value} \\
        \hline
        Learning rate & 1e-5 \\
        KL coefficient & 0.2 \\ 
        VF coefficient & 0.01 \\
        Clip-range & 0.05 \\
        Clip-range value & 0.2 \\
        Batch size & 8 \\
        Mini batch size & 4 \\
        Gradient accumulation steps & 4 \\
        Epoch & 1 \\
    \end{tabular}
    \label{tab:rl_hparams}
}
\end{table*}

\section{Experimental Results}\label{sec:exp_result}


\subsection*{\textbf{(RQ1) \rqone}}

       
        
        
        
        

        

\begin{table*}[h]
    \centering
    \caption{(RQ1) The performance of~\ours~when compared to state-of-the-art LLMs. LLMs without beam search are run 5 times and reported as Mean (SD). }
    \resizebox{\textwidth}{!}{
    \begin{tabular}{p{45pt}|l|p{50pt}|p{40pt}|p{75pt}|p{75pt}|p{75pt}|p{75pt}}
        \textbf{Evaluation Setting} & \textbf{Model} & \textbf{Model\break Parameters} & \textbf{Inference Time} & \textbf{Syntax\break Correctness}& \textbf{Passing \break Test Case} & \textbf{Code\break Coverage} & \textbf{Mutation \break Score} \\
        \hline
       
        \multirow{6}{*}{Raw} & \textbf{\ours~}& 770M & $<$1 hr & \textbf{99.42\%} & \textbf{84.41\%} & \textbf{80.98\%} & \textbf{61.45\%} \\
        
        & Finetuned CodeT5-large & 770M & $<$1 hr & 65.08\% & 35.56\% & 34.44\% & 26.94\% \\
        
        & InCoder & 6B & 250 hr & 4.63\% & 0.03\% & 0.03\% & 0.03\% \\
        
        & StarCoder & 15.5B & 40 hr & 96.78\% & 71.56\% & 69.07\% & 54.32\% \\
        
        & GPT3.5 & \textgreater175B & 12 hr & {93.59\% (0.79\%)} & {79.21\% (0.67\%)} & {75.10\% (0.68\%)} & {54.32\% (0.58\%)} \\
    
        
        & {Copilot} & {\textgreater12B} & {6 hr} & {59.40\% (0.87\%)} & {48.73\% (0.65\%)} & {47.11\% (0.60\%)} & {33.82\% (0.62\%)} \\

        \hline

        {\multirow{6}{*}{\shortstack[l]{3-test}}} & {\textbf{\ours~}} & {770M} & {$<$1 hr} & {\textbf{99.42\%}} & {\textbf{84.44\%}} & {\textbf{80.92\%}} & {\textbf{60.36\%}} \\

        & {Finetuned CodeT5-large} & {770M} & {$<$1 hr} & {96.01\%} & {53.57\%} & {51.58\%} & {40.23\%} \\

        & {InCoder} & {6B} & {250 hr} & {19.24\%} & {0.12\%} & {0.11\%} & {0.07\%} \\

        & {StarCoder} & {15.5B} & {40 hr} & {99.29\%} & {72.63\%} & {70.02\%} & {52.59\%} \\
        
        & {GPT3.5} & {\textgreater175B} & {12 hr} & {93.73\% (0.79\%)} & {80.27\% (0.78\%)} & {76.07\% (0.77\%)} & {54.53\% (0.60\%)} \\
    

        & {Copilot} & {\textgreater12B} & {6 hr} & {94.42\% (0.33\%)} & {72.47\% (0.22\%)} & {70.00\% (0.25\%)} & {53.18\% (0.29\%)} \\

        \hline

        {\multirow{6}{*}{\shortstack[l]{5-test}}} & {\textbf{\ours~}} & {770M} & {$<$1 hr} & {\textbf{99.42\%}} & {\textbf{84.41\%}} & {\textbf{80.91\%}} & {\textbf{60.41\%}} \\

        & {Finetuned CodeT5-large} & {770M} & {$<$1 hr} & {91.53\%} & {38.75\%} & {37.50\%} & {29.42\%} \\

        & {InCoder} & {6B} & {250 hr} & {12.83\%} & {0.03\%} & {0.03\%} & {0.03\%} \\

        & {StarCoder} & {15.5B} & {40 hr} & {98.96\%} & {71.86\%} & {69.34\%} & {54.38\%} \\
        
        & {GPT3.5} & {\textgreater175B} & {12 hr} & {93.72\% (0.79\%)} & {79.36\% (0.69\%)} & {75.23\% (0.68\%)} & {54.07\% (0.58\%)} \\
    

        & {Copilot} & {\textgreater12B} & {6 hr} & {90.72\% (0.53\%)} & {61.15\% (0.54\%)} & {59.12\% (0.55\%)} & {41.08\% (0.66\%)} \\

        \hline

        {\multirow{6}{*}{\shortstack[l]{Filtered by \\ Syntax \\ Correctness}}} & {\textbf{\ours~}} & {770M} & {$<$1 hr} & {99.63\%} & {\textbf{84.50\%}} & {\textbf{81.06\%}} & {\textbf{60.54\%}} \\

        & {Finetuned CodeT5-large} & {770M} & {$<$1 hr} & {98.93\%} & {38.39\%} & {36.90\%} & {24.86\%} \\

        & {InCoder} & {6B} & {250 hr} & {1.90\%} & {0.12\%} & {0.11\%} & {0.05\%} \\

        & {StarCoder} & {15.5B} & {40 hr} & {\textbf{99.97\%}} & {71.95\%} & {69.25\%} & {47.13\%} \\
        
        & {GPT3.5} & {\textgreater175B} & {12 hr} & {94.55\% (0.70\%)} & {79.79\% (0.72\%)} & {75.68\% (0.71\%)} & {54.76\% (0.59\%)} \\
    

        & {Copilot} & {\textgreater12B} & {6 hr} & {99.52\% (0.03\%)} & {59.59\% (0.95\%)} & {57.58\% (0.93\%)} & {39.56\% (0.83\%)} \\

        \hline
        
    \end{tabular}
    }
    \label{tab:rq1_sota}
\end{table*}




\underline{\emph{Results.}} \textbf{Of the test cases generated by \ours, 99\% are syntactically correct, 84\% are passed while achieving a code coverage of 80\% and a mutation score of 61\%, which outperforms all of the studied LLMs.}
Table~\ref{tab:rq1_sota} presents the performance of \ours~when compared to the state-of-the-art LLMs according to the four evaluation metrics.
When comparing \ours~with GPT3.5, our \ours~model is able to generate test cases that are 6\% more syntactically correct, 5\% more passing, 6\% more complete, and 7\% more effective in the shortest amount of inference time due to the smallest size of the model.
This finding indicates that a small language model like \ours~which is based on CodeT5 does not necessarily perform worse than large language models if they are carefully designed.
Instead, by incorporating the domain knowledge of software testing into the architecture of \ours~(i.e., the consideration of test case characteristics in the reward function that is largely ignored by existing LLM models), \ours~is able to perform better than large language models, with a smaller size of the model parameters and a shorter amount of inference time.
This highlights the significant novelty of our approach contributing to the research area of Text-to-Testcase generation.

\textbf{Our proposed DeepRL framework can improve the passing test case percentage by 49\%, the code coverage by 47\%, and the mutation score by 32\% when compared with the CodeT5 base model.}
Table~\ref{tab:rq1_sota} shows that, without the deep reinforcement learning, the CodeT5 base model performs worse than GPT3.5, which is aligned with the common intuition and prior studies~\citep{openai2023gpt4, guo2023exploring}.
On the other hand, when adding deep reinforcement learning to the CodeT5 base model, our experiment confirms that the \ours~model performs better than the GPT3.5 for all evaluation metrics.
This finding highlights the importance of our proposed deep reinforcement learning framework for Text-to-Testcase generation.

\begin{figure*}[t]
    \centering
    \includegraphics[width=0.95\textwidth]{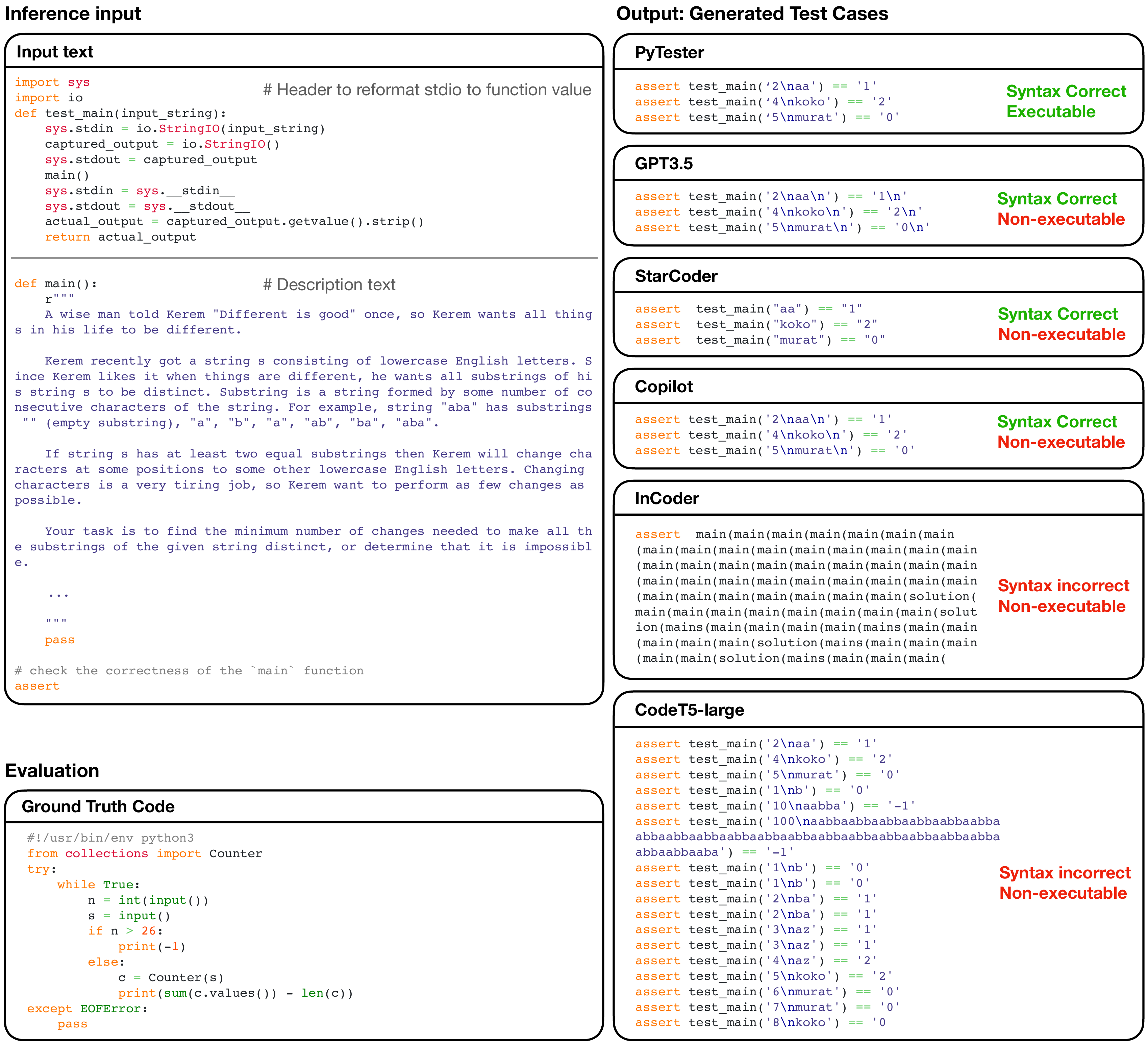}
    \caption{An example of test cases generated by \ours~and other baseline approaches (depicted from Problem ID \#569 of the testing dataset).
    }
    \label{fig:rq1_example}
\end{figure*}

Figure~\ref{fig:rq1_example} presents an illustrative example of the description input and the test cases by PyTester and the baselines.
This example depicts a programming task (Problem ID \#569) in the APPS dataset.
We observe that \ours~can generate test cases that are syntactically correct and align with the description while achieving high code coverage and high mutation score.
On the other hand, 

\begin{itemize}
    \item GPT3.5, Copilot and StarCoder can generate syntactically correct test cases, but some are non-executable, raising an \texttt{AssertionError}. 
    We observed that the non-executable test cases have to do with incorrect test inputs and incorrect expected outputs. 
    For example, StarCoder generates a test input with a missing of the first value (e.g., ``2\textbackslash n''), while Copilot and GPT3.5 generate the expected outputs of the test case with an additional ``\textbackslash n''.
    Such incorrectly generated test cases could be due to the lack of consideration of the passing test case during the learning process of the large language models.
    In contrast, the consideration of syntax correctness and code coverage in the reward function of \ours~improves the generated test cases to align to the given description (i.e., executable) with high coverage.
    \item InCoder often generates incorrect test cases (syntactically incorrect, non-executable, low code coverage, and low mutation score).
    Like in this example, some issues include the missing of an \texttt{assert} call, the incorrect test inputs/outputs, and the incompleted generation.
    Such incompleted generation happens since InCoder does not generate the terminal tokens (i.e., the \texttt{EOS} token), resulting in a longer inference time and incompleted assert statements.
    \item CodeT5-large often generates repetitive assert statements.
    For example, this assert statement ``\texttt{assert test\_main(`2\textbackslash nba') == `1'}'' is repeatedly generated two times, while the last assert statement is also incomplete.
    Such repetitive generation has to do with the traditional learning objective aiming to generate test cases that are exactly matched with the ground truth test cases.
    Differently, \ours~addresses this challenge by considering the test case characteristics during the test case generation.
    Rather than evaluating the test cases with a traditional text similarity measure, we instead consider the test case characteristics via a reward function.
\end{itemize}




\subsection*{\textbf{(RQ2) \rqtwo}}

\begin{table*}[h]
    \centering
    \caption{(RQ2) The performance of our \ours~for the 7 variations of the reward functions. \textbf{S}: Syntax, \textbf{E}: Executability, \textbf{C}: Coverage.}
    \label{tab:rq2_reward}
    \begin{tabular}{p{45pt}|p{50pt}|p{55pt}|p{45pt}|p{40pt}||p{40pt}}
    {\textbf{Reward\break Function}}    & {\textbf{Syntax\break Correctness}} & {\textbf{Passing \break Test Case}} & {\textbf{Code\break Coverage}} & {\textbf{Mutation Score}} & {\textbf{TestQuality}} \\
    \hline
    \textbf{S-only} & 99.39\%& 82.72\%      & 79.47\%     & 60.96\% &{69.00\%} \\
    \textbf{E-only} & 96.26\%& 75.73\% & 72.95\%     & 56.34\%   & 63.58\% \\ 
    \textbf{C-only} & 98.59\%& \textbf{86.87\%}  & \textbf{81.96\%}     & 56.47\%   & 66.87\% \\
    \textbf{S+E}                & 98.80\%& 83.49\%      & 80.34\%     & 61.25\% & {69.51\%} \\ 
    \textbf{S+C}& \textbf{99.42}\% & 84.41\% & 80.98\% & \textbf{61.45}\% & \textbf{69.87\%} \\ 
    \textbf{E+C}& 94.57\% & 67.78\% & 65.37\% & 51.01\% & 57.31\% \\ 
    \textbf{S+E+C} & 98.80\%& 81.68\% & 78.34\% & 58.76\% & 67.15\% 
    \end{tabular}
\end{table*}

\underline{\textit{Results.}}
\textbf{The top-3 reward functions for \ours~are the Syntax+Coverage, Syntax+Executability, and Syntax-only types of feedback.}
Table~\ref{tab:rq2_reward} presents the performance of our \ours~for the seven variations of the reward functions.
We find that the Syntax+Coverage, Syntax+Executability, and Syntax-only types of the reward function achieve a comparative TestQualityScore of 69.87\%, 69.51\%, and 69.00\%, respectively.
On the other hand, without considering the syntax correctness feedback (for the remaining reward function types), the TestQualityScore of \ours~is decreased by 3-13\%.
This finding highlights the importance of syntax correctness in the reward function---i.e., such feedback will greatly help the agent to generate test cases that are syntactically correct during the policy optimization step.
This finding suggests that syntax correctness must be the minimum consideration when designing a reward function for deep reinforcement learning in the text-to-test case generation task.

In addition, we also find that both Syntax+Coverage and Syntax+Executability mostly perform best (i.e., highly correlated).
For Syntax+Coverage, the model aims to directly generate syntactically correct test cases that achieve high code coverage, enabling the generated test cases to be executable against the ground truth code.
On the other hand, for Syntax+Executability, the model aims to directly generate syntactically correct test cases that are executable against the ground truth code, enabling the generated test cases to achieve high code coverage.
Therefore, considering either Syntax+Coverage or Syntax+Executability in the reward function is the best.
Nevertheless, the lowest-performing variation of our \ours~still outperforms most of the LLMs.

\subsection*{\textbf{(RQ3) \rqfour}}
\begin{figure*}[h]
    \centering
    \includegraphics[width=.9\textwidth]{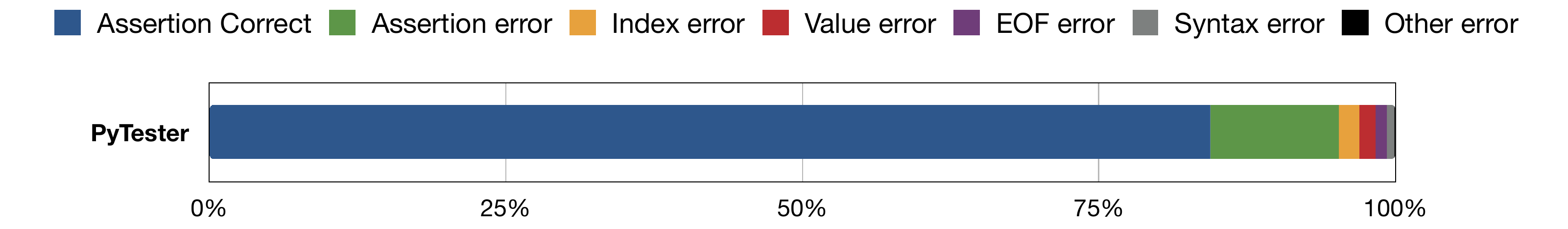}
    \caption{(RQ3) The common types of runtime errors that occur in the PyTester-generated test cases.}
    \label{fig:rq4_error}
\end{figure*}

\underline{\textit{Results.}} \textbf{Among the incorrectly generated test cases, \ours~mostly encounters an Assertion Error issue for 10.75\%.}
Despite \ours~performing the best when compared to the studied LLMs, we want to gain deeper insights into the incorrectly generated test cases.
As mentioned in Table~\ref{tab:rq1_sota}, 
we find that 84.41\% of the test cases generated from \ours~are passed
(the blue color in Figure \ref{fig:rq4_error}), meaning that 15.59\% of the generated test cases are not executable, thus incorrect.
Thus, we conduct an analysis to investigate what are the common types of runtime errors that occur in the PyTester-generated test cases.
As shown in Figure \ref{fig:rq4_error}, \ours~encounters the following top five common types of errors as follows:

\begin{itemize}
    \item Assertion Error (10.75\%) is an exception that is raised when an assert statement fails. The AssertionError is raised when the actual output (i.e., the output of the generated assertion) does not match the expected output (i.e., the output from the ground truth code). This finding highlights that as low as 10.75\% of the generated test cases are not passed the ground truth code.
    Therefore, future researchers should consider developing a novel mechanism integrated into the RL framework to improve the passing test case.
    \item Index Error (1.75\%) is an exception that occurs when trying to access an element from a sequence (such as a list, tuple, or string) using an index that is outside the valid range of indices for that sequence. For example, in Problem ID \#456, the model should generate the test case as ``\texttt{assert test\_main(`13\textbackslash n\colorbox{yellow}{o}gogmgogogogo') == `***gmg***'}'', but the model incorrectly generates the test case as ``\texttt{assert test\_main(`13\textbackslash ngogmgogogogo') == `***gmg***'}''.
    The highlighted character indicates the missing character that is not generated by \ours, leading to the incorrect access of the index in the sequence, thus, throwing an IndexError exception.
    \item Value Error (1.35\%) is an exception that occurs when a function or operation receives an input that has the correct data type but is not in the expected range or does not have a valid value. For example, in Problem ID \#911, the model should generate the test case as ``\texttt{assert test\_main(`\colorbox{yellow}{3 2}\textbackslash n50 85 250\textbackslash n10 15 25') == `Limak'}'', but the model incorrectly generates the test case as ``\texttt{assert test\_main(`\colorbox{yellow}{3}\textbackslash n2\textbackslash n1 1\textbackslash n2 3') == `2\textbackslash n0 points'}''.
    The Value Error is related to the highlighted character where it should receive two input values, but one of the values is missing, thus, throwing a ValueError exception.

    \item EOF Error (0.98\%) is an exception that occurs when a built-in function like \texttt{input()} or a method like \texttt{readline()} of a file object encounters an unexpected end of file (EOF) condition while trying to read input or data. For example, in Problem ID \#1152, the model should generate the test case as ``\texttt{test\_main(`3 3\textbackslash n0 1 0\textbackslash n0 1 0\textbackslash n1 0 0\textbackslash n1 0 0\textbackslash n1 0 0\textbackslash n\colorbox{yellow}{1 0 0}') == `Yes'}'', but the model incorrectly generated the test case as ``\texttt{test\_main(`3 3\textbackslash n0 1 0\textbackslash n0 1 0\textbackslash n1 0 0\textbackslash n1 0 0\textbackslash n1 0 0') == `Yes'}''. The EOF Error is related to one missing line (highlighted), thus, throwing an EOFError exception.
    
    \item Syntax Error (0.58\%) is an exception that occurs when the interpreter encounters a statement that does not conform to the rules of the Python programming language. For example, in Problem ID \#1184, the model incorrectly generates an incompleted assert statement as ``\texttt{assert test\_main(`[a, a, a, a, a, a, a}'', throwing a SyntaxError exception. However, such Syntax Error is relatively low when compared to other types of errors. This highlights the strengths of our approach that considers syntax correctness in the reward function, optimizing the models to generate more syntactically correct test cases.
\end{itemize}



\revise{R1.2, R1.3}{
\section{Discussion}\label{sec:discussion}

In this section, we discuss the additional evaluation of \ours~in the zero-shot setting on the HumanEval~\cite{chen2021humaneval} and MBPP~\cite{austin2021mbpp} datasets.
Notably, the HumanEval and MBPP datasets are relatively small, with sample sizes of $n$=164 and $n$=257, respectively. 


\textbf{On the HumanEval dataset, the raw output of \ours~ranks the second place in generating executable test cases with high code coverage, following only GPT-3.5.}
Table~\ref{tab:humaneval_sota} shows the results of \ours~compared to state-of-the-art LLMs and Pynguin on the HumanEval dataset.
Although performing in the zero-shot setting, the results show that \ours~can generate syntactically correct, executable test cases with high code coverage comparable to LLMs such as StarCoder and Copilot.
In fact, our \ours~model achieves the second rank in the raw evaluation, following only the GPT-3.5 model.
It is important to note that albeit providing the necessary information (i.e., description text, code, and test cases), the HumanEval dataset typically provides a \textit{concise docstring} for a function, different to the APPS dataset (i.e., the training dataset) that contains a \textit{lengthy, detailed description} for a function.
Therefore, this difference presents limitations that can impact the performance of \ours.

\textbf{On the MBPP dataset, \ours~performance is comparable to the StarCoder Model.}
Table~\ref{tab:mbpp_sota} shows the results of \ours~compared to state-of-the-art LLMs and Pynguin on the MBPP dataset.
Similar to the HumanEval dataset, the MBPP dataset has a unique nature of the text description, which is significantly different to the APPS dataset (i.e., the training dataset).
Specifically, the MBPP dataset provides only a \textit{single-line description} of a function, different to the APPS dataset that contains a \textit{lengthy, detailed description} for a function.
Therefore, this characteristic significantly impacts the performance of our \ours~that trained on a detailed description.
However, the results show that our \ours~can still perform better than the Finetune CodeT5-large and the InCoder models, and comparable if not better than the significantly larger model such as the StarCoder model.

\begin{table*}[h]
    \centering 
    \color{black}
    \caption{The performance of \ours~in the zero-shot setting on HumanEval when compared to state-of-the-art LLMs and Pynguin.}
    \resizebox{0.9\textwidth}{!}{
    \begin{tabular}    {p{45pt}|l|p{50pt}|p{75pt}|p{75pt}|p{75pt}|p{75pt}}
        \textbf{Evaluation Setting} & \textbf{Model} & \textbf{Model\break Parameters} & \textbf{Syntax\break Correctness}& \textbf{Passing \break Test Case} & \textbf{Code\break Coverage} & \textbf{Mutation\break Score} \\
        \hline
       
        \multirow{6}{*}{Raw} & \textbf{\ours~}& 770M & \textbf{99.39\%} & {83.54\%} & {80.09\%} & {48.74\%} \\
        
        & Finetuned CodeT5-large & 770M & 96.34\% & 68.90\% & 68.02\% & 49.80\% \\
        
        & InCoder & 6B & 12.20\% & 4.27\% & 4.16\% & 4.10\% \\
        
        & StarCoder & 15.5B & 85.37\% & 78.05\% & 77.25\% & 63.75\% \\
        
        & {GPT3.5} & {\textgreater175B} & \textbf{99.39\% (0.61\%)} & \textbf{89.51\% (0.80\%)} & \textbf{88.36\% (0.96\%)} & \textbf{69.59\% (0.76\%)} \\
    
        & {Copilot} & {\textgreater12B} & {83.78\% (0.92\%)} & {79.51\% (0.70\%)} & {79.00\% (0.69\%)} & {63.50\% (0.49\%)} \\

        & {Pynguin} & {-} & {86.59\%} & {83.54\% } & {82.87\%} & {25.59\%} \\

        \hline

        {\multirow{6}{*}{\shortstack[l]{3-test}}} & {\textbf{\ours~}} & {770M} & {{99.39\%}} & {{83.54\%}} & {{80.11\%}} & {{48.74\%}} \\

        & {Finetuned CodeT5-large} & {770M} & \textbf{100.00\%} & {73.17\%} & {71.67\%} & {50.58\%} \\

        & {InCoder} & {6B} & {54.88\%} & {14.63\%} & {14.28\%} & {11.80\%} \\

        & {StarCoder} & {15.5B} & \textbf{100.00\%} & {85.37\%} & {83.99\%} & {65.82\%} \\
        
        & {GPT3.5} & {\textgreater175B} & {99.51\% (0.51\%)} & \textbf{92.80\% (1.00\%)} & \textbf{91.19\% (1.27\%)} & \textbf{70.62\% (1.15\%)} \\
    
        & {Copilot} & {\textgreater12B} & {98.78\% (0.00\%)} & {89.63\% (0.43\%)} & {88.62\% (0.43\%)} & {70.52\% (0.25\%)} \\

        & {Pynguin} & {-} & {87.80\%} & {85.98\% } & {85.21\%} & {23.61\%} \\

        \hline

        {\multirow{6}{*}{\shortstack[l]{5-test}}} & {\textbf{\ours~}} & {770M} & {{99.39\%}} & {{83.54\%}} & {{80.11\%}} & {{48.74\%}} \\

        & {Finetuned CodeT5-large} & {770M} & {99.39\%} & {69.51\%} & {68.63\%} & {50.28\%} \\

        & {InCoder} & {6B} & {46.34\%} & {7.93\%} & {7.80\%} & {7.17\%} \\

        & {StarCoder} & {15.5B} & \textbf{100.00\%} & {80.49\%} & {79.65\%} & {64.35\%} \\
        
        & {GPT3.5} & {\textgreater175B} & {99.51\% (0.51\%)} & \textbf{89.87\% (1.26\%)} & \textbf{88.68\% (1.42\%)} & \textbf{69.62\% (0.89\%)} \\
    
        & {Copilot} & {\textgreater12B} & {97.32\% (0.33\%)} & {85.12\% (0.33\%)} & {84.51\% (0.33\%)} & {68.07\% (0.23\%)} \\

        & {Pynguin} & {-} & {85.37\%} & {84.15\% } & {83.43\%} & {21.45\%} \\

        \hline

        {\multirow{6}{*}{\shortstack[l]{Filtered by \\ Syntax \\ Correctness}}} & {\textbf{\ours~}} & {770M} & {{98.78\%}} & {{83.54\%}} & {{80.11\%}} & {{48.74\%}} \\

        & {Finetuned CodeT5-large} & {770M} & \textbf{100.00\%} & {68.90\%} & {68.04\%} & {49.80\%} \\

        & {InCoder} & {6B} & {35.37\%} & {4.27\%} & {4.16\%} & {4.10\%} \\

        & {StarCoder} & {15.5B} & \textbf{100.00\%} & {82.32\%} & {81.54\%} & {67.30\%} \\
        
        & {GPT3.5} & {\textgreater175B} & {99.76\% (0.33\%)} & \textbf{89.75\% (0.80\%)} & \textbf{88.60\% (0.98\%)} & \textbf{69.81\% (0.95\%)} \\
    
        & {Copilot} & {\textgreater12B} & {99.39\% (0.00\%)} & {83.54\% (0.61\%)} & {83.01\% (0.61\%)} & {66.50\% (0.52\%)} \\

        & {Pynguin} & {-} & {86.59\%} & {83.54\% } & {82.87\%} & {25.59\%} \\

        \hline
        
    \end{tabular}
    }
    \label{tab:humaneval_sota}

    \centering
    \color{black}
    \caption{The performance of \ours~in the zero-shot setting on MBPP when compared to state-of-the-art LLMs and Pynguin.}
    \resizebox{0.9\textwidth}{!}{
    \begin{tabular}{p{45pt}|l|p{50pt}|p{75pt}|p{75pt}|p{75pt}|p{75pt}}
        \textbf{Evaluation Setting} & \textbf{Model} & \textbf{Model\break Parameters} & \textbf{Syntax\break Correctness}& \textbf{Passing \break Test Case} & \textbf{Code\break Coverage} & \textbf{Mutation Score} \\
        \hline
       
        \multirow{6}{*}{Raw} & \textbf{\ours~}& 770M & {{90.66\%}} & {{15.18\%}} & {{14.95\%}} & {{5.74\%}} \\

        & {Finetuned CodeT5-large} & {770M} & {88.72\%} & {2.72\%} & {2.66\%} & {0.70\%} \\

        & {InCoder} & {6B} & {27.63\%} & {2.33\%} & {2.31\%} & {1.48\%} \\

        & {StarCoder} & {15.5B} & {43.19\%} & {9.73\%} & {9.62\%} & {5.20\%} \\
        
        & {GPT3.5} & {\textgreater175B} & \textbf{98.05\% (1.48\%)} & {37.04\% (1.54\%)} & {36.78\% (1.23\%)} & \textbf{21.97\% (1.30\%)} \\
    
        & {Copilot} & {\textgreater12B} & {61.56\% (1.39\%)} & {28.56\% (1.05\%)} & {28.54\% (1.05\%)} & {16.46\% (0.44\%)} \\

        & {Pynguin} & {-} & {89.49\%} & \textbf{79.77\% } & \textbf{75.86\%} & {20.71\%} \\
        
        \hline

        {\multirow{6}{*}{\shortstack[l]{3-test}}} & {\textbf{\ours~}} & {770M} & {{90.66\%}} & {{15.18\%}} & {{14.95\%}} & {{5.74\%}} \\

        & {Finetuned CodeT5-large} & {770M} & {98.44\%} & {4.67\%} & {4.61\%} & {1.87\%} \\

        & {InCoder} & {6B} & {61.87\%} & {2.72\%} & {2.69\%} & {1.48\%} \\

        & {StarCoder} & {15.5B} & {87.94\%} & {17.51\%} & {17.42\%} & {9.41\%} \\
        
        & {GPT3.5} & {\textgreater175B} & {98.44\% (1.46\%)} & {38.60\% (1.68\%)} & {38.24\% (1.50\%)} & \textbf{22.76\% (0.89\%)} \\
    
        & {Copilot} & {\textgreater12B} & \textbf{98.52\% (0.17\%)} & {38.68\% (0.59\%)} & {38.58\% (0.59\%)} & {21.61\% (0.43\%)} \\

        & {Pynguin} & {-} & {91.44\%} & \textbf{80.16\% } & \textbf{76.67\%} & {19.02\%} \\

        \hline

        {\multirow{6}{*}{\shortstack[l]{5-test}}} & {\textbf{\ours~}} & {770M} & {{90.66\%}} & {{15.18\%}} & {{14.95\%}} & {{5.74\%}} \\

        & {Finetuned CodeT5-large} & {770M} & {95.33\%} & {3.11\%} & {3.05\%} & {1.09\%} \\

        & {InCoder} & {6B} & {61.87\%} & {2.33\%} & {2.31\%} & {1.48\%} \\

        & {StarCoder} & {15.5B} & {78.99\%} & {12.84\%} & {12.73\%} & {6.57\%} \\
        
        & {GPT3.5} & {\textgreater175B} & \textbf{98.05\% (1.48\%)} & {37.04\% (1.54\%)} & {36.78\% (1.23\%)} & \textbf{21.97\% (1.20\%)} \\
    
        & {Copilot} & {\textgreater12B} & {95.85\% (0.39\%)} & {34.71\% (0.70\%)} & {34.67\% (0.70\%)} & {20.58\% (0.42\%)} \\

        & {Pynguin} & {-} & {90.66\%} & \textbf{80.16\% } & \textbf{77.00\%} & {21.71\%} \\

        \hline

        {\multirow{6}{*}{\shortstack[l]{Filtered by\\ Syntax \\ Correctness}}} & {\textbf{\ours~}} & {770M} & {{89.49\%}} & {{15.18\%}} & {{14.95\%}} & {{5.74\%}} \\

        & {Finetuned CodeT5-large} & {770M} & {99.22\%} & {2.72\%} & {2.66\%} & {0.70\%} \\

        & {InCoder} & {6B} & {32.30\%} & {2.33\%} & {2.31\%} & {1.48\%} \\

        & {StarCoder} & {15.5B} & {91.44\%} & {15.56\%} & {15.46\%} & {7.74\%} \\
        
        & {GPT3.5} & {\textgreater175B} & {98.36\% (0.51\%)} & {37.04\% (1.54\%)} & {36.78\% (1.23\%)} & \textbf{21.90\% (1.09\%)} \\
    
        & {Copilot} & {\textgreater12B} & \textbf{100.00\% (0.00\%)} & {35.25\% (0.65\%)} & {35.23\% (0.65\%)} & {19.80\% (0.28\%)} \\

        & {Pynguin} & {-} & {89.49\%} & \textbf{79.77\% } & \textbf{75.86\%} & {20.71\%} \\

        \hline
        
    \end{tabular}
    }
    \label{tab:mbpp_sota}
\end{table*}

}

\section{Related Works}\label{sec:related_work}

In this section, we highlight the difference in our work to the literature on reinforcement learning in software engineering.

Reinforcement learning (RL)~\citep{sutton2018reinforcement} is a machine-learning approach where an agent learns to map observations to actions by maximizing numerical reward functions.
Recently, RL has been applied to text generation tasks.
Specifically, LLMs utilize RL to train models based on user preferences through human feedback (RLHF)~\citep{christiano2017deep}. 
Consequently, LLMs can exhibit complex and coherent behaviors such as instruction following~\citep{ouyang2022training}, summarization~\citep{stiennon2020learning}, and conversation~\citep{jaques2019way}.

In software engineering, reinforcement learning has been adopted in the code generation tasks~\citep{wang2022compilable, le2022coderl} to improve the compilability and execution rate of the generated code.
For example, \citet{wang2022compilable} proposed CompCoder, an RL approach that considers compiler feedback in the reward function to generate more compilable code.
\citet{le2022coderl} proposed CodeRL, an actor-critic~\citep{bahdanau2016actor, konda1999actor} deep reinforcement learning framework for code generation. 

Different from prior studies, we formulate the Text-to-Testcase task as a deep reinforcement learning problem. 
We incorporate the domain knowledge such as test case characteristics (e.g., syntax and code coverage scores) into the model through a reward function. 
Additionally, we perform a deep analysis of ablation studies on reward components, showing the importance of the test case characteristics.
Finally, the results also show that our \ours~outperforms other LLMs, highlighting the novelty and the significance of our work that contributes to the area of Text-to-Testcase generation.


\section{Threat to Validity}\label{sec:threat}

In this section, we disclose our threat to the validity.

\textbf{Threats to construct validity.} The policy training and the policy optimization steps of our \ours~involve hyperparameter settings (see Table~\ref{tab:ft_hparams} and Table~\ref{tab:rl_hparams}).
Prior studies raised concerns that different hyperparameter settings may impact the performance of the Deep RL models~\citep{kiran2022hyperparameter}.
However, the parameter optimization is beyond the scope of this paper.
Thus, future research can consider exploring the impact of the hyperparameter settings on the Deep RL models.

Data leakage is often a critical concern when applying large language models (LLMs) in software engineering (i.e., the phenomenon when data samples in the testing dataset were included in the training dataset).
With PyTester, it is based on the CodeT5-large model, which is pre-trained on the CodeSearchNet dataset, and then later fine-tuned on the APPS benchmark dataset.
Prior to the pre-training and fine-tuning step, we carefully analyzed and ensured that none of the testing samples in the APPS dataset appeared in the CodeSearchNet~\citep{husain2019codesearchnet} and the training set of the APPS dataset to avoid any data leakage problem.
Thus, we confirm that the results of our PyTester will not be impacted by the data leakage.
With GPT-3.5 (baseline comparison), we are unable to determine if there is any data leakage during the model training of GPT-3.5 due to the close-source nature of GPT-3.5.
Despite the risk of data leakage with the GPT-3.5 model, our experiments still confirm that PyTester (with no risk of data leakage) outperforms GPT-3.5 (with the risk of data leakage) for all evaluation measures.


\textbf{Threats to internal validity.} The variation of the reward function may impact the performance of \ours.
However, our RQ2 confirms that, among the 7 variations, the performance varies between 57.31\% and 69.87\% of the TestCaseScore. 
Despite the high difference in the performance among the reward functions, we find that the lowest-performing variation of our \ours~still outperforms most of the LLMs.
Thus, the reward function does not pose a threat to the validity of our study.
Nevertheless, our study is limited by one scheme of reward points (ranging from -2 to 4).
However, finding an optimal scheme of reward points is beyond the scope of this work.
Thus, future research is encouraged to explore the impact of the reward points on the performance of Deep RL models.

The primary goal of software testing is to identify defects and vulnerabilities within the software. 
Consequently, automated test case generation is expected to produce test cases that can find bugs accurately. 
To evaluate this bug-finding capability of a model, a dataset containing buggy code is typically required.
However, due to the unavailability of the bug information in our APPS benchmark dataset, it is challenging to directly assess the bug-finding capability of PyTester.
To address this challenge, we include the evaluation of the mutation testing to our paper.
Mutation testing works by introducing small bugs into the correct code (e.g., replacing a minus operator with a plus operator). 
If the test cases still pass the mutated code, they will receive a zero Mutation Score, indicating that the test cases fail to identify bugs in the mutated code.
This metric provides valuable insight into how well and how sensitive the generated test cases can identify bugs, serving as an alternative means of assessing the potential capability of a model in bug-finding. 


\textbf{Threats to External Validity.} 
We acknowledge that our PyTester is currently limited to the generated test cases based on the textual description of one single call.
In real-world scenarios, test cases are more complex (including test prefixes, etc).
Thus, future research may consider the generation of more complex test cases that include test prefixes, etc.

\ours~training dataset is limited to the APPS benchmark dataset---the only Text-to-Testcase benchmark dataset with sufficient size.
In addition to broadening our analysis and ensuring the performance of \ours, we provide additional evaluations in the zero-shot setting on the HumanEval and MBPP datasets.
Future research may consider increasing the generalizability of our \ours~approach with other datasets, programming languages, and contexts.



\section{Conclusion}\label{sec:conclusion}

In conclusion, we propose \ours, a Text-to-Testcase generation approach for Python using Deep Reinforcement Learning via a reward function that considers three types of feedback, namely, syntax correctness, test executability, and code coverage.
Through a comprehensive experiment on the APPS benchmark dataset~\citep{hendrycks2021measuring}, we find that:

\begin{itemize}
    \item Of the test cases generated by \ours, 99\% are syntactically correct, 84\% are passed while achieving a code coverage of 80\% and a mutation score of 61\%, which outperforms all of the studied LLMs.
    \item  The top-3 reward functions for our PyTester are the Syntax+Coverage, Syntax+Executability, and Syntax-only types of feedback.
    \item As low as 15.59\% of the PyTester-generated test cases are incorrect. Among the incorrectly generated test cases, \ours~mostly encounters an Assertion Error issue for 10.75\%, followed by other types of errors (i.e., IndexError, ValueError, EOFError, SyntaxError).
\end{itemize}

Based on these results, we draw the following implications for future research.


\begin{itemize}
    \item \textbf{The performance of smaller language models is not necessarily inferior to that of large language models if carefully designed.}
    Training or finetuning large language models often requires significant computational resources, including powerful hardware and large datasets, with the assumption that more resources and more datasets will contribute to performance improvement~\citep{openai2023gpt4, guo2023exploring}.
    This paper takes a different direction from the existing work by using Deep RL to enable smaller language models, with the smallest model parameter size (770M) and the fastest inference times by \emph{at least} one order of magnitude, to outperform much larger language models (175B).
    This finding suggests that future research could consider improving small over large LMs for better resource efficiency.
    

    %




    \item \textbf{The SE domain knowledge should be considered when designing a deep reinforcement learning architecture.}
    The performance of many AI models often relies on its model architecture.
    Thus, many prior studies often adopted various data mining and optimization techniques to improve such AI models.
    Different from existing work, this paper takes an alternative direction by integrating the SE domain knowledge when designing the reinforcement learning architecture (i.e., incorporating test case characteristics into the reward function).
    This enables PyTester to generate test cases that are more syntactically correct, executable, complete, and effective while being aligned with a given description.
    

\end{itemize}



\bibliographystyle{elsarticle-harv}
\biboptions{authoryear}

\bibliography{mybibfile}


\end{document}